\documentclass[12pt,showpacs]{revtex4}
\usepackage[english]{babel}
\usepackage{graphicx}
\usepackage{epsfig}

\newcommand{\bea}{\begin{eqnarray}}
\newcommand{\eea}{\end{eqnarray}}
\newcommand{\be}{\begin{equation}}
\newcommand{\ee}{\end{equation}}
\newcommand{\sig}{\sigma}
\newcommand{\nn}{\nonumber}

\begin{document}

\title{
Extraction of proton form factors in the timelike region 
from single-polarized $e^+e^- \to \vec{p}\  \bar{p}$ events} 

\author{Andrea Bianconi}
\email{andrea.bianconi@bs.infn.it}
\affiliation{
Dipartimento di Chimica e Fisica per i Materiali e per l'Ingegneria,  
via Valotti 9, 25100 Brescia, Italy, and\\
Istituto Nazionale di Fisica Nucleare, Sezione di Pavia, I-27100 Pavia, Italy}

\author{Barbara Pasquini}
\email{barbara.pasquini@pv.infn.it}
  
\author{Marco Radici}
\email{marco.radici@pv.infn.it}
\affiliation{
Dipartimento di Fisica Nucleare e Teorica, Universit\`{a} di Pavia, and\\
Istituto Nazionale di Fisica Nucleare, Sezione di Pavia, I-27100 Pavia, Italy}

\begin{abstract}
We have performed numerical simulations of the single-polarized 
$e^+ e^- \to \vec{p}\  \bar{p}$ process in kinematic conditions under discussion for 
a possible upgrade of the existing DAFNE facility. By fitting the cross section and
spin asymmetry angular distributions with typical Born expressions, 
we can study the conditions for extracting
information on moduli and phases of the proton electromagnetic form factors in the 
timelike region, which are poorly known and whose preliminary data show 
puzzling features. We have explored also non-Born contributions by introducing a 
further component in the angular fit, which is related to two-photon exchange 
diagrams. Using a dipole parametrization, we show that these corrections can be 
identified if larger than 5\% of the Born contribution; we also explore the conditions
for extracting information on the phase and, consequently, on the relative weight
between their real and imaginary parts, which are presently unknown.
\end{abstract}


\maketitle

\section{Introduction}
\label{sec:intro}

The electromagnetic form factors are one of the most relevant 
sources of information on the structure of hadrons and on their internal 
quark/gluon dynamics. A lot of data for
nucleons have been accumulated in the spacelike region using elastic electron 
scattering (for a review, see Ref.~\cite{gao} and references therein). While the 
traditional Rosenbluth separation method suggests the well known scaling of the 
ratio $G_E/G_M$ of the electric to the magnetic Sachs form factor, new 
measurements on the electron-to-proton polarization transfer in 
$\vec{e}^- p \to e^- \vec{p}$ scattering reveal contradicting results, 
with a monotonically decreasing ratio for increasing momentum transfer 
$-q^2=Q^2$~\cite{jlab}. 
This has led to several theoretical works in order to test the reliability of the
Born approximation underlying the Rosenbluth method (see Ref.~\cite{2gamma} and
references therein). 

Hadron form factors are experimentally less well known  
in the timelike region $q^2$ $>$ 0, where 
they can be explored in $e^+ e^-$ annihilations or hadron-hadron collisions 
(for a review see Ref.~\cite{egle1}). Being complex objects, their absolute values 
can be extracted by combining the measurement of total cross sections and 
center-of-mass (c.m.) angular distributions of the final products. The phases are 
related to the polarization of the involved hadrons 
(see, e.g., Refs.~\cite{egle1,dubnick, brodsky1}), but they have not yet been 
measured. The available unpolarized differential cross sections 
have always been integrated over a wide 
angular range, such that the relative weight of $\vert G_M\vert$ and 
$\vert G_E\vert$ is still unknown. In the data analysis either hypothesis $G_E=0$ 
or $|G_E|=|G_M|$ were used for all $q^2$, which are both not justified {\em a
priori}.  

Nevertheless, the few available results show quite interesting properties.
The amplitudes in the timelike and spacelike regions are connected 
by dispersion relations~\cite{dr}; consequently, at sufficiently  
large $\vert q^2\vert$ the form factors should 
behave in a similar way in both regions. A fit to 
the existing proton $|G_M|$ data for $q^2 \leq 20$ GeV$^2$
~\cite{e685-2} suggests that surprisingly this asymptotic regime has not yet 
been reached. Moreover, 
the very recent data from the BaBar collaboration on $|G_E/G_M|$~\cite{babar} 
show that the ratio is larger than 1, contradicting the 
spacelike results with the polarization transfer method~\cite{jlab} and the 
previous timelike data from LEAR~\cite{lear}. Also the few neutron $|G_M|$ data
are unexpectedly larger than the proton ones in the corresponding $q^2$ 
range~\cite{fenice}. Finally, very close to the threshold 
$q^2 = 4m^2$ (with $m$ the nucleon mass), the trend of data suggests that the form 
factors could be affected by interesting
subthreshold resonance structures (for more details, see Ref.~\cite{baldini}). 

The ongoing discussion about the upgrade of the DAFNE facility~\cite{LoI} by
enlarging the c.m. energy range from the $\phi$ mass to 2.5 GeV while keeping a
luminosity of $10^{32}$ cm$^{-2}$s$^{-1}$, and, in particular, by inserting a
polarimeter around the interaction region~\cite{roadmap}, would allow to explore the
production of polarized $p\bar{p},\  n\bar{n},\  \Lambda\bar{\Lambda},$ and  
$\Sigma\bar{\Sigma}$ pairs with great precision. As for protons, in a previous
paper~\cite{noi-unpol} we explored the conditions under which the ratio $|G_E/G_M|$
could be extracted at DAFNE-2 from the angular distribution of the unpolarized cross
section at any given $q^2$. We showed that with $300\,000$ events in the considered
$3.8\leq q^2\leq 6.2$ GeV$^2$ region, the uncertainty on the ratio is about
10\%. Here, we consider the case where the final proton is polarized normally to the
reaction plane. Using the same sample of events, we try to determine the conditions 
to realistically extract the phase of $G_E/G_M$ from the simulation of a single-spin 
asymmetry measurement.

We also made some simulations of non-Born contributions, mainly related to two-photon
$(2\gamma)$ exchange diagrams, which seem to play a crucial role in the analysis of 
the $G_E/G_M$ ratio in the spacelike region~\cite{jlab}. Theoretically, 
these mechanisms are very poorly known; in particular, the relative size of the real 
and imaginary parts is undetermined. Therefore, in our simulations we explore the 
most favourable conditions to set upper limits on their magnitude and verify if their 
contribution can be measured at DAFNE-2. In Ref.~\cite{noi-unpol}, we assumed the 
$2\gamma$ amplitude mainly real, and we came to the preliminary conclusion that 
their effect can be isolated only if their absolute size is at least 5\% of the Born 
contribution. Here, at variance, we check that if the $2\gamma$ amplitude is 
mainly imaginary, its effect could be detected more easily in single-spin asymmetries 
than from unpolarized cross sections.

In Sec.~\ref{sec:formulae}, we briefly review the necessary general formalism. In
Sec.~\ref{sec:mc}, we outline the main features of our Monte Carlo simulation of the
$e^+ e^- \to \vec{p}\  \bar{p}$ process. In Sec.~\ref{sec:reconstr}, we describe the
steps to extract information on form factors by reconstructing the simulated normal
polarization with a two-parameter fit. In Sec.~\ref{sec:out}, we discuss the results.
Finally, some concluding remarks are given in Sec.~\ref{sec:end}


\section{General formalism} 
\label{sec:formulae}

The scattering amplitude for the reaction $e^+ e^- \to p \bar{p}$, 
where an electron and
a positron with momenta $k_1$ and $k_2,$ respectively, 
annihilate into a proton-antiproton pair 
with momenta $p_1$ and $p_2,$ respectively, 
is related by crossing to the
corresponding scattering amplitude for elastic $e^- p$ scattering. 
There are several
equivalent representations of it; 
here, we use the one involving the axial 
current following the scheme of Ref.~\cite{egle2}. The scattering amplitude
can be fully parametrized in terms of 
three complex form factors: $G_E(q^2,t), G_M(q^2,t),$ and $G_A(q^2,t)$, 
which are 
functions of $q^2=(k_1+k_2)^2$ and $t=(k_2-p_1)^2$. They refer to an 
identified proton-antiproton pair in the final state, rather than to an 
identified isospin state. 
In the following, we will use $\cos\theta$ instead of the variable $t$, 
with $\theta$ the angle between 
the momenta of the positron and of the recoil proton in the c.m. frame. 

In the Born approximation, 
$G_E$ and $G_M$ reduce to the usual Sachs form factors and do not depend on 
$\cos\theta$, while $G_A=0$. 
For the construction of our Monte Carlo event generator, we rely on 
the general and extensive 
relations of Ref.~\cite{egle2} up to the single polarization case. 
These relations assume small non-Born terms 
and include them up to order $\alpha^3$ (with $\alpha$ the fine structure 
constant), i.e. they consider $2\gamma$ exchanges only via their interference 
with the Born amplitude. These corrections introduce explicitly six new functions
that all can depend on $\cos\theta$:  
the real and imaginary parts of $\Delta G_E$ and $\Delta G_M$, i.e. of the 
$2\gamma$ corrections to the Born magnetic and electric form factors, and the 
real and imaginary parts of the axial form factor $G_A$. 

The role of these 
corrections in the unpolarized  and single-polarized
cross sections is very simple:
it can be deduced from the Born term
by adding the  contribution of $G_A$ and
substituting the Born form factors with the 
"$2\gamma-$improved" ones, i.e.   $G_{E,M}(q^2)$ $\rightarrow$ 
$G_{E,M}(q^2)+\Delta G_{E,M}(q^2,cos\theta).$ 
In Ref.~\cite{egle2}, this fact 
is somehow hidden by neglecting terms of order $\alpha^4$.
For sake of simplicity, we keep $2\gamma$ effects via the axial form factor only. 
Within this scheme, when summing events with positive and 
negative polarization, the unpolarized cross section can be written as
\bea
\frac{d\sig^o}{d\cos\theta} &= 
& a(q^2) \, [ 1+ R(q^2) \, \cos^2 \theta ] - b(q^2) \, \mathrm{Re}
[G_M(q^2)\,{G_A}^\ast(q^2,\cos \theta)]\, \cos \theta \; , \label{eq:unpolxsect} 
\\
a(q^2) &= &\frac{\alpha^2 \pi}{2q^2}\,\frac{1}{\tau}\,\sqrt{1-\frac{1}{\tau}}\,
\left( \tau |G_M|^2 + |G_E|^2\right) \; , \quad 
b(q^2) = \frac{2\pi \alpha^2}{q^2}\,\frac{\tau -1}{\tau} \; ,\label{eq:ab} \\
R(q^2) &= &\frac{\tau |G_M(q^2)|^2-|G_E(q^2)|^2}{\tau |G_M(q^2)|^2+|G_E(q^2)|^2} 
\; , \quad \tau = \frac{q^2}{4m^2} \; . \label{eq:rtau}
\eea

When the proton is polarized, the cross section is
linear in the spin, i.e. $d\sig = d\sig^o \, (1+ \mathcal{P} \, 
\mathcal{A} )$, with $d\sig^o$ from Eq.~(\ref{eq:unpolxsect}) and
$\mathcal{A}$ the analyzing power. In the c.m. frame, three 
polarization states are observable~\cite{dubnick,brodsky1}: the longitudinal 
$\mathcal{P}_z$, the sideways $\mathcal{P}_x$, and the normal $\mathcal{P}_y$. 
The first two ones lie in the scattering plane, while the normal points in the 
$\mathbf{p}_1 \times \mathbf{k}_2$ direction, the $x, y, z,$ forming a 
right-handed coordinate system with the longitudinal $z$ direction along
the momentum of the outgoing proton. The $\mathcal{P}_y$ is particularly 
interesting, since it is the only observable that does not require a polarization 
in the initial state~\cite{dubnick,brodsky1}. With the above approximations, it can 
be deduced by the spin asymmetry between events with positive and negative 
perpendicular polarizations:
\bea
\mathcal{P}_y &= &\frac{1}{\mathcal{A}_y}\, 
\frac{d\sig^\uparrow - d\sig^\downarrow}{d\sig^\uparrow + d\sig^\downarrow} 
\nn \\
&=
&\frac{b(q^2)}{2\sqrt{\tau -1}\, d\sig^o}\, \sin \theta \, 
\Big\{ \cos \theta \, \mathrm{Im}\left[ G_M(q^2) \, G_E^\ast(q^2) \right]
\nn\\
& & - 
\sqrt{\frac{\tau -1}{\tau}}\, \mathrm{Im}\left[ G_E(q^2) \, 
{G_A}^\ast(q^2,\cos \theta) 
\right] \Big\} \; .
\label{eq:py}
\eea
This spin asymmetry is produced by the mechanism $\mathbf{p}_1 \times \mathbf{k}_2
\cdot \mathbf{S}_B$, which is forbidden in the Born approximation
for the spacelike elastic scattering~\cite{brodsky1}.
 
Measurements of the unpolarized distribution~(\ref{eq:unpolxsect}) and spin 
asymmetry~(\ref{eq:py}) at fixed $q^2$ for different $\theta$ allow to fit the 
different angular terms, from which we may extract absolute values and relative
phases of $G_E/G_M$, and some information on $G_A(q^2,\cos\theta)$. The Born
contributions to $d\sigma^o$ and $\mathcal{P}_y$ have a typical $\cos^2\theta$ and 
$\sin 2\theta$ behaviours,
respectively, any deviation due to non-Born terms. The normal $\mathcal{P}_y$
vanishes at end points $\theta = 0,\pi$ and at threshold $q^2=4m^2 (\tau = 1)$.
Interestingly, at $\theta = \pi/2$ the Born contribution vanishes, and 
$\mathcal{P}_y$ gives direct insight to the $2\gamma$ amplitude~\cite{egle2}. 

However, we note that the measurement of 
$\mathcal{P}_y$ alone does not completely determine the phase 
difference of the 
complex form factors. By defining with $\delta_E$ and $\delta_M$ the phases of the 
electric and magnetic form factors, respectively,
the Born contribution is proportional to 
$\sin (\delta_M - \delta_E)$, leaving the ambiguity between 
$(\delta_M - \delta_E)$ and $\pi - (\delta_M - \delta_E)$. Only
the further measurement of $\mathcal{P}_x$ can solve the problem, because 
$\mathcal{P}_x \propto \mbox{\rm Re}(G_M\,G_E^\ast) \propto \cos (\delta_M -
\delta_E)$~\cite{brodsky1}. But at the price of requiring a polarized electron 
beam.


\section{General features of the numerical simulations}
\label{sec:mc}

We consider the $e^+ e^- \to \vec{p}\  \bar{p}$ process. Most of the details of the
simulation are mutuated from a previous work~\cite{noi-unpol}. Therefore, events are 
generated in the usual variables $q^2, \, \theta ,\, \phi$ (the azimuthal angle of the 
proton momentum with respect to the scattering plane), and $S_y$ (the proton 
polarization normal to the scattering plane). Then, the distribution is integrated 
upon $\phi$ but not summed upon the spin. 

The exchanged timelike $q^2$ is fixed by the beam energy. Only the scattering 
angle $\theta$ is randomly distributed. For a given $q^2$, the Born terms in the
unpolarized cross section~(\ref{eq:unpolxsect}) and spin asymmetry~(\ref{eq:py}) are  
responsible for the $[1+R(q^2)\,\cos^2 \theta ]$ and 
$\sin 2\theta \, \mathrm{Im}[ G_M(q^2) \, G_E^\ast(q^2)]$ behaviours, respectively. 
Since with the discussed approximations no other dependence in $\theta$ is present, 
observed systematic deviations from these behaviours will be interpreted as a clear 
signature of non-Born terms. The present lack of knowledge on such mechanisms forces us
to further simplify the picture and to neglect the $\cos\theta$ dependence of the 
axial form factor by taking the first term of its expansion in powers of $\cos\theta$
(see Ref.~\cite{noi-unpol} for further details).

An overall sample of $300\,000$ events has been 
considered with $3.8 \leq q^2 \leq 6.2$ GeV$^2$ and $\vert \cos\theta \vert < 0.9$. 
We recall that the integrated cross section for $e^+ e^- \to p \bar{p}$ in the 
considered region is approximately 1 nb; hence, at the foreseen luminosity of 
$10^{32}$ cm$^{-2}$s$^{-1}$ for DAFNE-2~\cite{LoI,roadmap} this sample can be 
collected in one month with efficiency 1. The lower $q^2$ bin does not include the 
$p\bar{p}$ threshold, because this region is characterized by peculiar mechanisms 
like, e.g., the Coulomb focussing and possible subthreshold resonances. As it will be
evident later, the spin asymmetry between events with positive and negative $S_y$
polarization is very small for the first $q^2$ bin and would suggest to exclude it from
the analysis. However, this would imply to reduce the size of the sample by
almost a factor 2. In fact, the event distribution approximately falls like
$1/q^{10}$~\cite{brodsky,noi-unpol}, leaving the bins at higher $q^2$ scarcely 
populated. For more details, we refer to the discussion in Ref.~\cite{noi-unpol}. Here,
we just remark that the interesting quantities related to the spin
asymmetry~(\ref{eq:py}) increase approximately in a linear way from threshold to $q^2
\sim 5-6$ GeV$^2$: while absolute errors increase with $q^2$ because of less populated 
bins, the relative errors do not (see discussion in Sec.~\ref{sec:out}). Contrary to the 
unpolarized case, there is no reason 
to asymmetrically split the beam time in order to make the statistics of each $q^2$ 
bin more homogeneous. In any case, in the following we will use the "default
conditions", as they were defined in Ref.~\cite{noi-unpol}: the $q^2$ range is divided
in 6 equally spaced bins with width $\Delta q^2 = 0.4$ GeV$^2$; for each of
them, the solid angle $\vert \cos\theta\vert < 0.9$ is divided in 7 equally spaced 
bins with width $\Delta \cos\theta \approx 0.257$. For each of the 42 bins, events are 
further divided according to their polarization, and the spin 
asymmetry is constructed according to Eq.~(\ref{eq:py}) by taking the ratio between the
difference and the sum of events with positive or negative polarization.

Events can be generated only by inserting specific parametrizations of the proton
form factors in the cross section. For $G_E$ and $G_M$, several models can be
considered in the timelike region~\cite{brodsky1,egle1}, mostly derived from 
extrapolations from the spacelike region. As in Ref.~\cite{noi-unpol}, we select the 
parametrizations of Refs.~\cite{iachello,lomon}, because they have been recently 
updated in Ref.~\cite{egle1} by simultaneously fitting both the spacelike and timelike 
available data. Moreover, both cases release separate parametrizations for the
real and imaginary parts of $G_E$ and $G_M$, as they are needed in
Eqs.~(\ref{eq:unpolxsect}) and (\ref{eq:py}). We indicate the former as the IJLW
parametrization (from the initials of their authors), and the latter as the Lomon
parametrization. 

Since the explored $q^2$ range is not large, it is reasonable to make also much
simpler, but equally effective, choices. In the previous work~\cite{noi-unpol}, we 
used parametrizations where all form factors are real and proportional to the 
same dipole term $1 / (1 + q^2/q_o^2)^2$, with $q_o^2 = 0.71$ GeV$^2$, as in the spacelike 
region. The actual parameters were reduced to the ratios $r_e = \vert G_E/G_M \vert$ 
and $r_a = \vert G_A/G_M\vert$. However, this choice leads to a vanishing spin
asymmetry, since all form factors have the same phase. Here, we generalize 
the case aiming to alternatively exphasize each one of the two contributions in 
Eq.~(\ref{eq:py}), and analyze the conditions for their extraction. We define the
parametrization Dip1$i$ as the one with $r_e=1,\  r_a=0$, and where the relative 
phase of $G_E$ with respect to $G_M$ is $\beta_e = \pi/2$, i.e. both $G_M$ and $G_E$ 
have a dipole trend in $q^2$ but the former is real while the latter is purely 
imaginary; there are no effects from $2\gamma$ exchanges because $G_A=0$. With 
Dip1$i$, only the Born term survives in Eq.~(\ref{eq:py}) and its effect is 
maximized. The second choice is named Dip$2\gamma i$ and it is defined by 
$r_e = 1,\ r_a=0.2$, and $\beta_e = 0,\  \beta_a = \pi/2$, with $\beta_a$ the 
relative phase of $G_A$ with respect to $G_M$. Namely, $G_E$ and $G_M$ are equal and 
real; $G_A$ has the same dipole trend, but it is purely imaginary and with a modulus 
5 times smaller than $|G_M|$. In this case, in Eq.~(\ref{eq:py}) only the non-Born 
term is active and is maximized. 

The Dip1$i$ parametrization must not be considered realistic but rather an exercise
to set the upper limit on $\mathrm{Im}[G_E]$. In fact, for the considered $q^2$ range 
it is reasonable to expect $|G_E| \sim |G_M|$, since the two moduli are equal at 
threshold. Moreover, $\mathrm{Im}[G_E]$ cannot be large considering that the 
absorption induced by the $p-\bar{p}$ rescatterings is much below the 
unitarity limit (see Ref.~\cite{noi-unpol}, and references therein). Indeed, the Born
contribution to the asymmetry induced by the Lomon parametrization, turns out to be
not larger than 10\%; at these $q^2$, the IJLW parametrization gives even a vanishing
asymmetry. On the other side, we notice that nothing can be presently said about the 
relative weight of $\mathrm{Re}[G_A]$ and $\mathrm{Im}[G_A]$. If for the considered 
$q^2$ range we can approximately take $\mathrm{Im}[G_{E/M}] \sim 0$, we deduce from 
Eq.~(\ref{eq:unpolxsect}) that the $\mathrm{Re}[G_A]$ can be best extracted from the 
unpolarized reaction and it was analyzed in Ref.~\cite{noi-unpol}. Viceversa, the 
assumption in Dip2$\gamma i$ of a large $\mathrm{Im}[G_A]$ selects the normal 
polarization of Eq.~(\ref{eq:py}) as the best observable to study $2\gamma$ 
exchanges. In this sense, the choice Dip$2\gamma i$ is also an attempt to set a 
possible upper limit to non-Born contributions. However, as in Ref.~\cite{noi-unpol} 
we stress that we set $r_a = |G_A|/|G_M| = 0.2$ to account for these effects as 
corrections to the Born result. In particular, we note that a correction of 20\% does 
not contradict the experimental findings on $G_E/G_M$ in the spacelike region with 
the polarization transfer method~\cite{jlab}.  


\begin{figure}[h]
\centering
\includegraphics[width=8cm]{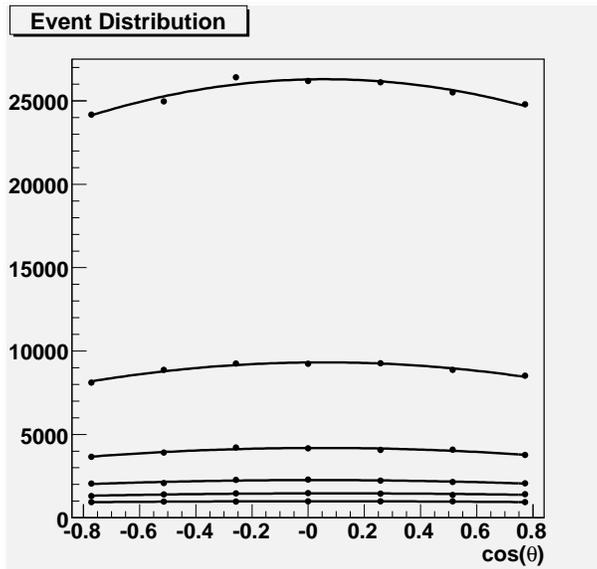}   
\caption{Angular distribution of $300\,000$ events for the $e^+ e^- \to p\bar{p}$
process at $3.8\leq q^2 \leq 6.2$ GeV$^2$ and $\vert \cos\theta \vert < 0.9$, and
with the Lomon parametrization of proton form factors (see text). The points
correspond to 7 equally spaced $\cos\theta$ bins for each of 6 equally spaced
$q^2$ bins. Solid curves are the results of a 3 parameters angular fit with
$B\propto r_a =\vert G_A/G_M\vert = 0$ (see text). The
highest curve corresponds to the lowest bin $3.8 \leq q^2 \leq 4.2$ GeV$^2$; next
lower curve to the adjacent bin $4.2 \leq q^2 \leq 4.6$ GeV$^2$, and so on.
\label{fig:distr_lomon}}
\end{figure}


\section{Reconstruction of the spin asymmetry} 
\label{sec:reconstr}

In Ref.~\cite{noi-unpol}, we analyzed in detail the conditions for an
unpolarized $e^+ e^- \to p\bar{p}$ measurement at DAFNE-2~\cite{LoI,roadmap} in order
to estimate the achievable precision in the reconstruction of $r_e = |G_E/G_M|$ and in
a possible detection of $2\gamma$ contributions. As it has been recalled in
Sec.~\ref{sec:intro}, the complete determination of $G_E$ and $G_M$ in the timelike
region requires the knowledge of their relative phase, which in turn involves the
polarized $e^+ e^- \to \vec{p}\  \bar{p}$ measurement.

Our analysis consists of five steps. For sake of simplicity, we describe it in
the following for the case of the Lomon parametrization:
\begin{itemize}
\item[1)] We generate $300\,000$ events for $3.8 \leq q^2 \leq 6.2$ GeV$^2$ and 
$\vert \cos\theta\vert < 0.9$ (i.e., for 
$25^{\mathrm{o}} \leq \theta \leq 155^{\mathrm{o}}$). The sorted events are summed upon
the positive and negative polarizations $S_y$ and are 
divided into 6 equally spaced $q^2$ bins of width $\Delta q^2 = 0.4$ GeV$^2$. In 
each $q^2$ bin, the events are divided into 7 equally spaced $\cos\theta$ bins 
with width $\Delta \cos\theta \approx 0.257$. In Fig.~\ref{fig:distr_lomon}, the points
show the 6 unpolarized distributions in $q^2$, each one consisting of 7 points 
describing the distribution in $\cos\theta$. From the discussion in the previous 
section, the highest $\cos\theta$ distribution corresponds to the lowest bin 
$3.8 \leq q^2 \leq 4.2$ GeV$^2$; the next lower distribution to the adjacent bin
$4.2 \leq q^2 \leq 4.6$ GeV$^2$, and so on.

\item[2)] For each $q^2$ bin, the $\cos\theta$ dependence of the unpolarized event 
distribution is fitted by a function of the form 
\be
C_q(\cos\theta) \equiv A\, N_q(\cos\theta) =\ A\Big(1\ +\ R\, \cos^2\theta\ -\ B \, 
\cos\theta\Big)\;, 
\label{eq:dsig0fit}
\ee
where the three fitting parameters $A, R$, and $B$, change for each different $q^2$ 
bin. The results of each fit are represented by the solid curves in 
Fig.~\ref{fig:distr_lomon}. The parameter $A$ is related to the total number of collected
events and it does not play any role in the following analysis; $R$ is 
connected to the angular coefficient $R(q^2)$ of Eq.~(\ref{eq:rtau}), and $B$ to the 
correction from $2\gamma$ exchanges (for a thorough discussion, see 
Ref.~\cite{noi-unpol}). 

\begin{figure}[h]
\centering
\includegraphics[width=8cm]{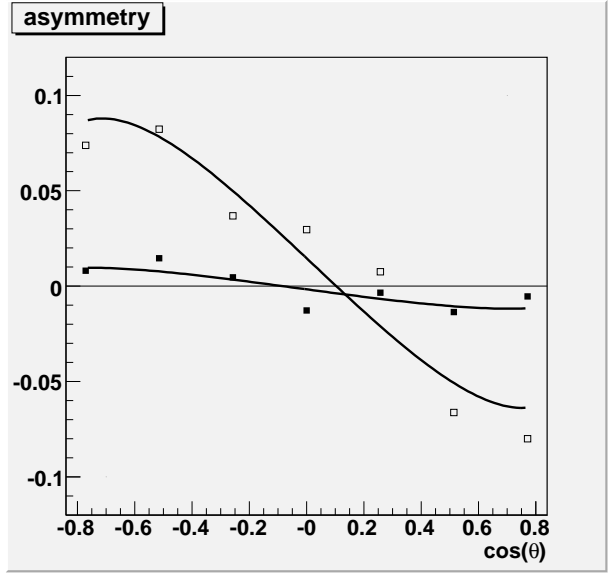}   
\caption{Normal polarization for the $e^+ e^- \to \vec{p}\  \bar{p}$ process in the
same conditions as in Fig.~\ref{fig:distr_lomon}. Full squares refer to the lowest bin 
$3.8 \leq q^2 \leq 4.2$ GeV$^2$, empty squares to the highest $5.8 \leq q^2 \leq 6.2$ 
GeV$^2$. Solid lines are the result of a 2 parameter angular fit (see text).
\label{fig:asym_lomon}}
\end{figure}

\item[3)] For each $q^2, \theta$ bin, the events generated at step 1) are divided in 
two groups corresponding to positive $(U)$ and negative $(D)$ normal polarizations 
$S_y$. Then, the spin asymmetry is manually constructed by taking the ratio 
$(U-D) / (U+D)$. In Fig.~\ref{fig:asym_lomon}, the full squares show the $\cos\theta$
distribution for the lowest $q^2$ bin, while the empty squares refer to the largest
$q^2$ bin.

\item[4)] After fixing $N_q(\cos\theta)$, for each $q^2$ bin the $\theta$ 
dependence of the spin asymmetry is fitted by a function of the form
\be
A_q(\theta) = \frac{\sin\theta}{N_q(\cos\theta)}\, \left( A_1\, \cos\theta - A_2
\right) \; ,
\label{eq:pyfit}
\ee
where the fit parameters $A_1$ and $A_2$ refer, respectively, to the Born term
$\mathrm{Im}[G_M^{} G_E^\ast]$ and to the non-Born term $\mathrm{Im}[G_E^{} G_A^\ast]$
in Eq.~(\ref{eq:py}). The results of the fits for the lowest and largest $q^2$ bins are
represented by the solid lines in Fig.~\ref{fig:asym_lomon}. 


\begin{figure}[h]
\centering
\includegraphics[width=8cm]{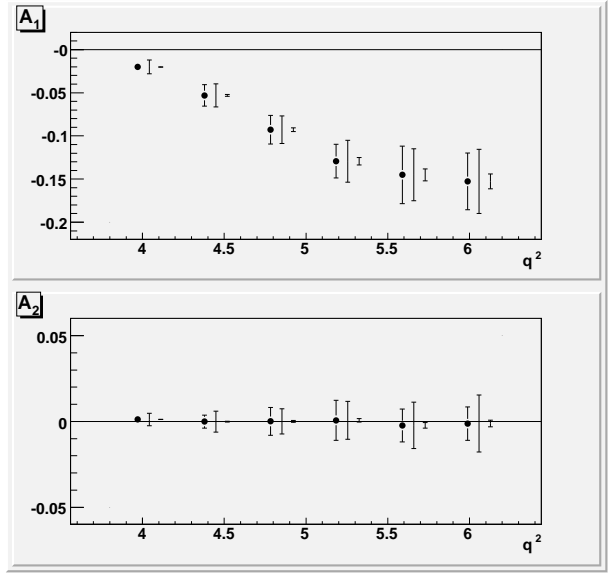}   
\caption{Values of the fitting parameters $A_1$ and $A_2$ extracted from the normal
polarization of Fig.~\ref{fig:asym_lomon} according to Eq.~(\ref{eq:pyfit}). For each 
point, the leftmost error bar gives the statistical error, the central error bar gives 
the fitting (systematic) error, the rightmost error bar is the uncertainty associated 
with the fitting error on the unpolarized event distribution. 
\label{fig:final_lomon}}
\end{figure}

\item[5)] The procedure is repeated 10 times, starting from statistically independent
samples, and average values and statistical variances for $A_1$ and $A_2$ are built for
each $q^2$ bin and displayed in Fig.~\ref{fig:final_lomon}. We checked that stability
of the results is ensured already after 6 repetitions. In the figure, each point is 
accompanied also by two other error bars, whose meaning is the following. For each
repetition, $A_1$ and $A_2$ are determined with some error from the fitting procedure.
From left to right, the second error bar associated to each point is the average of the
fitting errors upon the 10 repetitions. We recall that this error gives the integral
deviation of the test function from the data points; therefore, it must be considered
as a systematic error and, in principle, it has nothing to do with the statistical 
one. Finally, the rightmost error bar associated to each point gives the 
uncertainty on $A_1$ and $A_2$ deriving from the uncertainty on the separately fitted 
parameters $R$ and $B$: when the latter ones oscillate between the extreme values 
allowed by their fitting error bars, the averages values of $A_1$ and $A_2$ can 
oscillate inside the rightmost error bars displayed in Fig.~\ref{fig:final_lomon}. 
\end{itemize}

The Lomon parametrization includes only Born contributions to the complex $G_E$ and 
$G_M$, i.e. $G_A=0$ in Eqs.~(\ref{eq:unpolxsect}) and (\ref{eq:py}). Consequently,
the unpolarized angular distributions of Fig.~\ref{fig:distr_lomon} are symmetric 
functions in $\cos\theta$ and the spin asymmetry of Fig.~\ref{fig:asym_lomon} displays 
the typical $\sin2\theta$ trend, as it should be. Consistently, the fit parameter
$A_2$, related to the interference between $G_A$ and $G_E$, turns out to vanish in
Fig.~\ref{fig:final_lomon} within the statistical (and fitting) uncertainty. 

The fits of the unpolarized distribution and of the spin asymmetry are separately 
performed, because each single generated event can be collected in the group with
positive $(U)$ or negative $(D)$ normal polarization, defining the asymmetry as $(U-D)
/ (U+D)$. Our procedure corresponds to separately fit the denominator $U+D$ (with the
3 fit parameters $A, R$, and $B$) and the numerator $U-D$ (with the 2 fit parameters
$A_1$ and $A_2$). The uncertainty in the former fitting procedure does not affect much
the latter one. This can be visualized by the rightmost error bar associated with
each point in Fig.~\ref{fig:final_lomon}. We checked that this is not accidental by
repeating the fit of the spin asymmetry but artificially assuming the denominator $U+D$
independent from $\cos\theta$: the results on $A_1$ and $A_2$ stay basically the same.
It must be noted, however, that in the considered $q^2$ range both the angular
coefficient $R(q^2)$ of Eq.~(\ref{eq:rtau}) and the non-Born contribution proportional
to $\mathrm{Im}[G_M^{} G_A^\ast]$ are small, because $|G_E| \approx |G_M|$ and $G_A$ is
assumed to vanish in the Lomon parametrization. In general, for larger $q^2$ the first
condition may not be satisfied, while the second one appears too drastic. Consequently,
the analysis of error bars would become more involved; but a discussion along this line
is beyond the scope of this paper.

As previously described, the first and second leftmost error bars in
Fig.~\ref{fig:final_lomon} represent the statistical and fitting errors on $A_1$ and
$A_2$. The fitting error is usually a systematic error, except in those situations 
where in the limit of a large number of events the test function reproduces exactly 
the event distribution. This may happen when the number of fit parameters is equal to 
the number of bins (which is not our case, because we have 2 parameters against 7 bins
in $\cos\theta$), or when the test function is the correct guess for the event
distribution. In general, this is not possible since, in reality, the $\theta$ 
dependence of the form factors is unknown. But this is the case of our simulation,
where we have neglected the $\theta$ dependence in all form factors and the fitting
functions~(\ref{eq:dsig0fit}) and (\ref{eq:pyfit}) reproduce the complete angular
dependence of the event distributions. Therefore, in the limit of a large number of
events the statistical and fitting errors become equal; for the $300\,000$ events
considered here, they are very similar, as it is evident from
Fig.~\ref{fig:final_lomon}. The main consequence is that, in the conditions previously 
discussed, the statistical and fitting errors are two different ways to estimate 
the same error, and they must not be summed as one usually does for statistical and
systematic errors. This statement can be safely generalized when the analysis is
performed in the Born approximation, because $G_E$ and $G_M$ depend only on $q^2$ and
the fitting functions~(\ref{eq:dsig0fit}) and (\ref{eq:pyfit}) with $B=0$ and $A_2=0$,
respectively, are certainly correct. It is more questionable when non-Born
contributions are included, either in $G_{E/M}$ or via $G_A$, and the $\cos\theta$
dependence should be considered also inside the form factors.


\begin{figure}[ht]
\centering
\includegraphics[width=8cm]{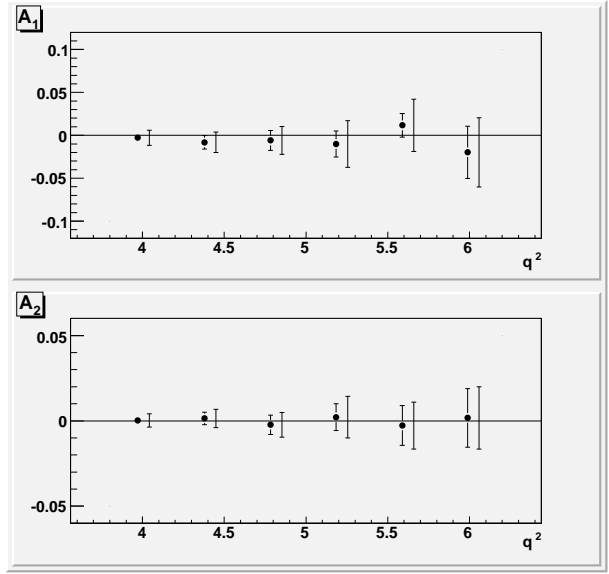}
\caption{The same as in Fig.~\ref{fig:final_lomon}, but for the IJLW parametrization 
(the rightmost error bar for each point is not displayed). 
\label{fig:final_IJLW}}
\end{figure}


\section{Discussion of results}
\label{sec:out}

In the following, we discuss the output of repeating the procedure outlined in the
previous section but for different parametrizations, as they have been described in
Sec.~\ref{sec:mc}. For sake of brevity, we will omit the figures with the angular 
distributions and will show only the dependence on $q^2$ of the parameters $A_1$ and 
$A_2$ extracted from the fit. For each point, only the statistical and fitting errors 
will be displayed, since the sensitivity to the uncertainty on the $R$ and $B$ parameters
of Eq.~(\ref{eq:dsig0fit}) is small (see Fig.~\ref{fig:final_lomon}). 

In Fig.~\ref{fig:final_IJLW}, the average values of $A_1$ and $A_2$ with statistical
and fitting errors are displayed for the considered $q^2$ bins for the IJLW
parametrization. The $A_2$ is consistently compatible with zero because no
$2\gamma$ effects are included. Surprisingly, the $A_1$ is also compatible with zero, 
while at much larger $q^2$ this choice gives very large asymmetries, similarly to the 
Lomon parametrization~\cite{egle1}. Luckily, the relatively small error bars allow to
unambigously distinguish the two realistic parametrizations, except maybe for the
lowest $q^2$ bin. 

It is not accidental that the cleanest comparison is possible for the largest $q^2$
here considered. At the threshold $q^2_{th} = 4m^2$, we have $G_E = G_M$ and the Born 
term in Eq.~(\ref{eq:py}) vanishes. For $q^2$ not much larger than $q^2_{th}$, we may
assume that 
\be
G_E \approx G_M \, \left[ 1 - \alpha \  (q^2-q^2_{th}) + \ldots \right]\, 
e^{i \left[ \beta \  (q^2-q^2_{th})+\ldots \right]} \; .
\label{eq:linear}
\ee
The Born term of the spin asymmetry will behave like $-\sin[\beta (q^2-q^2_{th})]\, 
[1 - \alpha (q^2-q^2_{th})] \approx -\beta (q^2-q^2_{th})$ for small $q^2-q^2_{th}$.
Hence, it is easy to justify the approximate linear dependence for $q^2\leq 5$ 
GeV$^2$ displayed both in Fig.~\ref{fig:final_lomon} and Fig.~\ref{fig:final_IJLW}. 

\begin{figure}[ht]
\centering
\includegraphics[width=8cm]{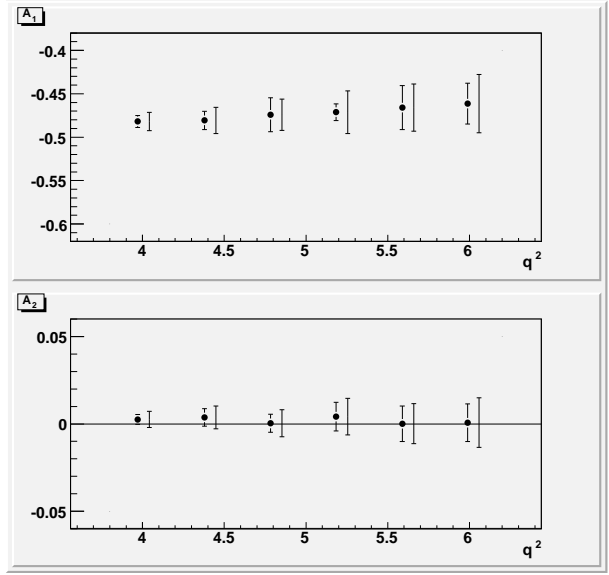}   
\caption{The same as in Fig.~\ref{fig:final_lomon}, but for the Dip1$i$ 
parametrization (the rightmost error bar for each point is not displayed).
\label{fig:final_dip1i}}
\end{figure}

As anticipated in Sec.~\ref{sec:mc}, we have also considered the Dip1$i$
parametrization which emphasizes the Born contribution to $\mathcal{P}_y$ by taking
$G_A=0$ and $G_E$ purely imaginary with the same modulus as $G_M$. The corresponding
fit parameters $A_1$ and $A_2$ are displayed in Fig.~\ref{fig:final_dip1i}: $A_2$ is
obviously consistent with zero, while $A_1$ is roughly three times larger in size than
in the more realistic parametrizations Lomon and ILJW. Evidently, it must be considered 
as an upper limit.

\begin{figure}[ht]
\centering
\includegraphics[width=8cm]{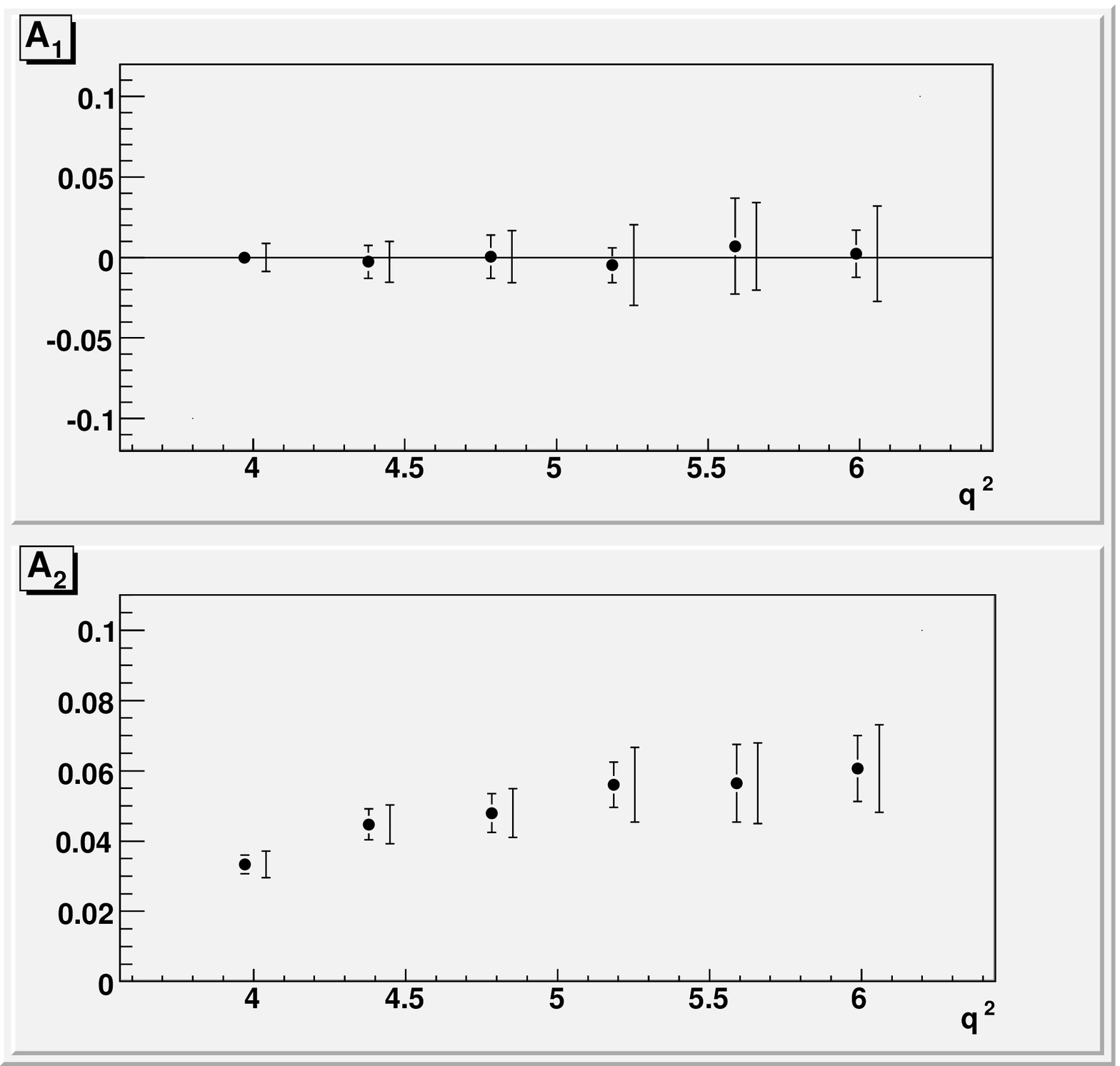}   
\caption{The same as in Fig.~\ref{fig:final_lomon}, but for the Dip$2\gamma i$ 
parametrization (the rightmost error bar for each point is not displayed).
\label{fig:final_dip2gi}}
\end{figure}

Similarly, the parametrization Dip$2\gamma i$ emphasizes the
non-Born contribution in $\mathcal{P}_y$ by taking $G_E$ and $G_M$ real and $G_A$
purely imaginary. As already discussed in Sec.~\ref{sec:mc}, this is not an upper limit
for the $2\gamma$ effects, since we have imposed $|G_A|/|G_M| = 0.2$, consistently with
the findings in the spacelike region about the $G_E/G_M$ ratio~\cite{jlab}. But it
represents the most favourable situation to explore $\mathrm{Im}[G_A]$ using the spin 
asymmetry. In Fig.~\ref{fig:final_dip2gi}, $A_1$ is obviously consistent with zero, and 
$A_2$ indicates a clearly detectable effect. 

Analogously to the Born case, the $2\gamma$ effects seem better visible at
the largest $q^2$ here explored, even if the latter correspond to the least
populated bins. We note also that while the Born term in $\mathcal{P}_y$ approximately
falls like $|G_E/G_M|$ for increasing $q^2$ and, therefore, is not expected to fall
fastly unless far beyond 10 GeV$^2$~\cite{babar}, viceversa, from Eq.~(\ref{eq:py})
together with Eqs.~(\ref{eq:unpolxsect},\ref{eq:ab}), the weight of the $2\gamma$ 
contribution should fall at least as $1/\sqrt{\tau} \sim 1/q$. Therefore, unless the 
$2\gamma$ effect increases with $q^2$, the available range of its measurability is 
reduced to $q^2 < 10$ GeV$^2$.


\section{Conclusions}
\label{sec:end}

We have performed numerical simulations of the single-polarized 
$e^+ e^- \to \vec{p}\  \bar{p}$ process using a sample of $300\,000$ events in the
kinematical region with $3.8\leq q^2 \leq 6.2$ GeV$^2$, distributed over 6 equally
spaced bins. For each $q^2$ bin, event have been further distributed over 7 equally 
spaced bins in $\cos(\theta)$, with $|\cos\theta|<0.9$. For each $q^2, \theta$ bin, 
the events have been separated in two groups according to their positive $(U)$ or
negative $(D)$ normal polarization of the recoil proton with respect to the reaction
plane. 

The angular distributions of $U+D$ (the unpolarized cross section) has been fitted
with a three-parameter function of the type $A\, [1+R\  \cos^2\theta - B\  \cos\theta]$,
where from the parameters $R$ and $B$ information on $r_e=|G_E/G_M|$ and
$r_a=|G_A/G_M|$ can be deduced, respectively. The angular distribution $U-D$
(proportional to the spin asymmetry) has been fitted with the two-parameter function
$\sin\theta\, [A_1\  \cos\theta - A_2]$, where the two free parameters $A_1$ and $A_2$
are related to $\mathrm{Im}[G_M^{} G_E^\ast]$ and $\mathrm{Im}[G_E^{} G_A^\ast]$,
respectively, i.e. to the relative phases of the complex form factors. 

Using realistic parametrizations for $G_E$ and $G_M$ deduced by fitting the available
spacelike and timelike data, the statistical uncertainty makes it possible to 
unambiguously distinguish among different models, which are equivalent in the
spacelike domain. Therefore, both moduli and phases can be extracted with a two-step
fitting procedure, where the fitting errors of the first step do not affect much 
the uncertainty in the second step. Surprisingly, the extraction of phases seems
better performed for the higher $q^2$ bins, because the spin asymmetry is small for
$q^2$ close to the threshold and the relative error becomes smaller and smaller for
increasing $q^2$.

The additional contribution related to two-photon exchange diagrams shows up as a
deviation from the expected angular trend of the Born contribution, via the parameters
$B$ and $A_2$. In our analysis, we modeled such contribution in terms of the complex
axial form factor $G_A$, constraining its modulus not to contradict the findings in
the spacelike region. Nothing can be said about the relative weight of
$\mathrm{Re}[G_A]$ and $\mathrm{Im}[G_A]$. However, assuming that for the considered
$q^2$ range $G_E$ and $G_M$ are similar and with small imaginary parts, information on
$\mathrm{Re}[G_A]$ and $\mathrm{Im}[G_A]$ can be extracted from the non-Born term of
the unpolarized cross section and of the spin asymmetry, respectively. Using a simple
dipole parametrization, our finding is that the non-Born term can be identified if 
its size is at least 5\% of the Born contribution; 
if $\mathrm{Im}[G_A]$ is the dominant term, the asymmetry related to the $A_2$ coefficient can be clearly
measured, again with a preference for the highest $q^2$ bins.

\section*{Acknowledgements}
The work of B.P. and M.R. is partially supported  by the EU Integrated
Infrastructure Initiative Hadron Physics under contract number
RII3-CT-2004-506078.



\end{document}